\documentclass{SLS2014}

\usepackage{natbib}
\usepackage{proof}
\usepackage{amssymb}
\usepackage{amsmath}
\usepackage{mathrsfs}
\usepackage{mathbbol}
\usepackage{enumitem}
\usepackage{thmtools}
\usepackage{thm-restate}
\usepackage{hyperref}
\hypersetup{
 colorlinks,
     citecolor = {black},
    linkcolor={black},
    urlcolor={black}
}
\usepackage[official]{eurosym}
\DeclareRobustCommand{\officialeuro}{%
  \ifmmode\expandafter\text\fi
  {\fontencoding{U}\fontfamily{eurosym}\selectfont e}}

\usepackage{tikz-cd}

\usepackage{lineno}

\renewcommand{\phi}{\varphi}

\makeatletter
\def\labelandtag#1#2{\begingroup
   \def\@currentlabel{#2}%
   \phantomsection\label{#1}\endgroup
}
\makeatother

\newcommand{\myhypertarget}[2]{%
  \phantomsection
  \hypertarget{#1}{#2}%
  \expandafter\gdef\csname targettext@#1\endcsname{#2}%
}
\newcommand{\myhyperlink}[1]{%
  \hyperlink{#1}{\csname targettext@#1\endcsname}%
}

\newcommand{\base}[1]{\mathscr{#1}}
    \newcommand{\baseB}{\base{B}}
    \newcommand{\baseC}{\base{C}}

\newcommand{\baseGeq}{\supseteq}

\newcommand{\suppM}[2]{\Vdash_{#1 }^{ #2 }}

\newcommand{\At}{\mathbb{A}}

\DeclareSymbolFont{bbsymbol}{U}{bbold}{m}{n}
\DeclareMathSymbol{\fatsemi}{\mathbin}{bbsymbol}{"3B}
\DeclareMathSymbol{\fatcomma}{\mathbin}{bbsymbol}{"2C}

\newcommand{\GenRes}{\mathbb{R}}

\makeatletter
\let\orgdescriptionlabel\descriptionlabel
\renewcommand*{\descriptionlabel}[1]{%
  \let\orglabel\label
  \let\label\@gobble
  \phantomsection
  \edef\@currentlabel{#1}%
  \let\label\orglabel
  \orgdescriptionlabel{#1}%
}
\makeatother

\usepackage{setspace}
\title{A Note on an Inferentialist Approach to Resource Semantics 
}

\author{
{\small Alexander V. Gheorghiu$^{1}$ \qquad Tao Gu$^{1}$ \qquad David J. Pym$^{1,2,3}$} \\[1.5ex] { 
\small \{alexander.gheorghiu.19, tao.gu.18, d.pym\}@ucl.ac.uk}
}

\begin{document}

\maketitle \vspace{-7mm}

\footnotetext[1]{ Department of Computer Science, University College London, London WC1E 6BT,~UK}

\footnotetext[2]{ Department of Philosophy, University College London, London WC1E 6BT,~UK}

\footnotetext[3]{ Institute of Philosophy, University of London, London WC1E 7HU, UK
}

\makeatletter
\let \thefootnote = 4
\makeatother

A central concept within informatics is that of a \emph{distributed system}~\citep{CDKB-DS-2011}. From the systems modelling perspective, with a little abstraction, one can think of a distributed system as comprising delimited collections of interconnected component systems ~\citep{CMP2012,CP15}. These systems ultimately comprise interconnected \emph{locations}, at which are situated \emph{resources}, relative to which \emph{processes} execute --- consuming, creating, moving, combining, and otherwise manipulating resources as they evolve, so delivering the system's services. This broad definition applies across the many complex socio-economic-technical systems that define and support our world including physical and organizational infrastructure (e.g., city roads,  management structures, etc.), telecommunication networks (e.g., the world wide web), and real-time process control systems (e.g., aircraft control systems).

An important use of logic is to represent, understand, and reason about such systems; this determines the field of \emph{logical systems modelling}. To this end, one requires an interpretation of logical formulae in terms of the resources and states of the system; such an interpretation is called a \emph{resource semantics} of the logic. We propose the following encompassing definition: \vspace{-2mm}
\begin{quote} \small 
    \emph{A resource semantics for a system of logic is an interpretation of its formulae as assertions about states of processes and is expressed in terms of the resources that are manipulated by those processes}.  
\end{quote}\vspace{-2mm}
There is no intended restriction on the assertions that are to be included in the scope of this definition and assertions may refer not only to ground states but may also be `higher-order' statements about state transitions. Moreover, the \emph{manipulation} of resources by the system's processes is general and includes, \emph{inter alia}, consuming/using resources, creating resources, copying/deleting resources, and moving resources between locations.

There are, perhaps, two canonical examples of resource interpretations of logics that merit discussion, the  \emph{number-of uses} reading of \emph{Linear Logic} (LL)~\citep{girard1987linear} and the \emph{sharing/sepa\-ra\-tion} reading of \emph{the logic of Bunched Implications} (BI)~\citep{o1999logic}. While both are applicable to systems modelling, these two readings work in different ways and are related to different ways of doing the modelling. The number-of-uses reading ~\citep{Lafont,Hoare} for LL concerns the dynamics of the system: formulae denote processes and resources themselves. Note, it is available only through LL's proof theory, and is not reflected in LL's truth-functional semantics \citep{Girard}. 
 Meanwhile, the sharing/separation reading ~\citep{pym2019resource} of BI concerns the composition and state of the system; this is best exemplified by its essential place in \emph{program verification} --- see, for example, the use of \emph{Separation Logic}~\citep{reynolds2002separation}. It proceeds through the (ordered-monoidal/relational) model-theoretic semantics of BI.
 
Both readings are useful, but they are individually limited for 
systems modelling: sharing/separation expresses the \emph{structure} of distributed systems, and number-of-uses expresses the \emph{dynamics} of the resources involved. 
Also, they operate in completely different paradigms: number-of-uses proceeds from LL's proof theory, while sharing/separation proceeds from the semantics of BI. 
We suggest that \emph{proof-theoretic semantics} (P-tS)~\citep{SEP-PtS} offers a unified framework for resource semantics able to express both parts of a distributed system.

In the semantic paradigm of P-tS, 
\emph{inferentialism},  meaning emerges from rules of inference ~\citep{Brandom2000}. This stands in contrast to, for example, \emph{denotationalism}, wherein the meaning of logical structures is grounded in reference to abstract algebraic structures, as in 
model-theoretic semantics. 
Much of the groundwork for P-tS has been provided by \cite{Prawitz2006natural}, \cite{Dummett1991logical}, \cite{Schroeder2006validity}, and \cite{Piecha2019incompleteness}. 

Recently, \cite{GGP2023-IMLL} and \cite{Yll-arxiv} have provided P-tS for (intuitionistic) LL, and \cite{GGP2024-BI} have provided P-tS for BI, all building on essential work by \cite{Sandqvist2015base}. Relative to these results, \cite{GGP2024-IRS} have initiated an inferentialist account of resource semantics that uniformly expresses both number-of-uses readings and sharing/separation interpretations of the logics. Moreover, they have illustrated its application to both modelling the security architecture of an airport and to multi-factor authentication.  

These semantics are all given by an inductively defined judgment relation called \emph{support} such that transmission of validity 
is defined as follows: 
\begin{equation}{  
\label{eq:general-inf} 
    \Gamma \suppM{\baseB}{S(\cdot)} \phi 
\quad \text{iff} \quad
    \forall \baseC \baseGeq \baseB, \forall U \in \GenRes(\At), \text{ if } \suppM{\baseC}{U} \Gamma, \text{ then } \suppM{\baseC}{S(U)} \phi 
    \tag{Gen-Inf}
}
\end{equation} 
Let us spell out each ingredient of \eqref{eq:general-inf} and their roles in the resource semantics, which describes the execution of a certain policy in a system:
\begin{itemize}[label=--]
    \item $\phi$ specifies an assertion describing (a possible state of) the system
    \item $\Gamma$ specifies a policy describing the 
    executions of a system's processes 
    \item $\GenRes(\At)$ is some concept of atomic resource
    \item $S(\cdot)$ is some contextual atomic resource --- that is, whenever $S(\cdot)$ is combined with some atomic resource in $U \in \GenRes(\At)$, it returns a `richer' atomic resource $S(U)$
    \item $\baseB, \baseC$ are `bases' (a basic concept in P-tS) and \emph{model} the policy governing the system described by the assertions. 
\end{itemize}
Putting these all together, the support judgement $\Gamma \suppM{\baseB}{S(\cdot)} \phi$ says that, if policy $\Gamma$ were to be executed with contextual resource $S(\cdot)$ based on the model $\baseB$, then the result state would satisfy $\phi$. \eqref{eq:general-inf} explains how such execution is triggered: for arbitrary model $\baseC$ that extends $\baseB$, if policy $\Gamma$ is met in model $\baseC$ using some resource $U$, then $\Gamma$ could be executed, and the resulting state --- which consumes $U$ in the available contextual resource $S(\cdot)$ --- satisfies $\phi$. This summarizes an inferentialist approach to resource semantics. 

Distributed systems play a crucial role in society 
and logic is a primary tool for representing, understanding, and reasoning about them, with resource semantics being a key tool. Adopting logical inferentialism, particularly through recent advancements in the proof-theoretic semantics of substructural logics used to handle the two primary approaches to resource semantics, offers a unified approach for handling resource semantics in the logical modelling of systems. This approach not only encapsulates existing methodologies but also introduces novel, refined features, enhancing our capabilities in this domain. 


{
}

\end{document}